\documentclass[twoside]{article}
\usepackage{qic,epsfig}

\usepackage{amsmath}
\usepackage{multirow}
\usepackage{cite}

\newcommand{\myeta}{r}

\begin{document}
\setlength{\textheight}{8.0truein}    


\normalsize\textlineskip
\thispagestyle{empty}
\setcounter{page}{1}


\vspace*{0.88truein}

\alphfootnote

\fpage{1}

\centerline{\bf
TIME-SHIFT ATTACK IN PRACTICAL QUANTUM CRYPTOSYSTEMS
}
\vspace*{0.37truein}
\centerline{\footnotesize
BING QI}
\vspace*{10pt}
\centerline{\footnotesize
CHI-HANG FRED FUNG}
\vspace*{10pt}
\centerline{\footnotesize
HOI-KWONG LO}
\vspace*{10pt}
\centerline{\footnotesize
XIONGFENG MA}
\vspace*{10pt}
\centerline{\footnotesize\it
Center for Quantum Information and Quantum Control (CQIQC),
}
\baselineskip=10pt
\centerline{\footnotesize\it
Dept. of Physics and Dept. of Electrical and Computer Engineering,
}
\baselineskip=10pt
\centerline{\footnotesize\it
University of Toronto, Toronto, Ontario M5S 3G4, Canada
}
\vspace*{0.225truein}

\vspace*{0.21truein}

\abstracts{ Recently, a new type of attack, which exploits the
efficiency mismatch of two single photon detectors (SPD) in a
quantum key distribution (QKD) system, has been proposed. In this
paper, we propose another ``time-shift'' attack that exploits the
same imperfection. In our attack, Eve shifts the arrival time of
either the signal pulse or the synchronization pulse or both between
Alice and Bob. In particular, in a QKD system where Bob employs
time-multiplexing technique to detect both bit ``0'' and bit ``1''
with the same SPD, Eve, in some circumstances, could acquire full
information on the final key without introducing any error.
In addition, we prove that if Alice and Bob are unaware of our attack,
the final key they share is insecure.
We emphasize that our attack is simple and feasible with current technology.
Finally,
we discuss some counter measures against our and earlier attacks.
}{}{}

\vspace*{10pt}

\vspace*{3pt}

\vspace*{1pt}\textlineskip    









\section{Introduction\label{introduction}}

One of the most important practical application of quantum information is
quantum key distribution (QKD), whose unconditional security is
based on the fundamental law of quantum mechanics
\cite{Bennett1984,Ekert1991,Gisin2002,Mayers2001,Lo1999,Shor2000}.
In principle, any eavesdropping attempt by a third party (Eve) will
unavoidably introduce disturbance. So, it's possible for the
legitimate users (Alice and Bob) to upper bound the amount of
information acquired by an eavesdropper from the measured quantum
bit error rate (QBER). Alice and Bob can then distill a final
secure key by performing error correction and privacy amplification.
Unfortunately, a practical QKD system has imperfections and Eve may try
to exploit these imperfections and launch specific attacks. One
example is the photon number splitting (PNS)
attack\cite{PNS1,PNS2,PNS3,PNS4}, where Eve takes advantage of the
nonzero multi-photon emission probability of Alice's laser source
and the lossy nature of a quantum channel. Security proofs for
practical QKD systems exist \cite{Inamori2005,Gottesman2004}.
Recently, decoy state QKD
\cite{Hwang2003,Lo2004,Lo2005,Wang2005a,Ma2005b,Wang2005b,Harrington,Yi2005}
has also been proposed
to substantially increase
the distance and key generation rate of a practical QKD system when weak coherent-state sources are used.

Another critical device, which limits the performance of a long
distance QKD system, is the single photon detector (SPD). Currently,
many practical QKD systems use optical communication fibers as their
quantum channels and work at the Telecom wavelengths of around
either 1550nm or 1310nm. Single photon detection at these
wavelengths is often performed by (thermoelectric) cooled InGaAs
avalanche photo diodes (APDs) \cite{Ribordy2004}. To minimize the
dark count rate, this type of detector usually works in gated
mode: it is activated by a suitable electrical gate signal for a
narrow window (a few ns) only when a signal pulse (a few hundreds
of ps) is expected to arrive. Outside of this time window, a detector has no
response to the input photons. Obviously, in this design, the
synchronization between Alice's source and Bob's detectors is
critical.

Recently, an eavesdropping attack that exploits the efficiency
mismatch of two single photon detectors in a practical QKD system
has been proposed in \cite{Makarov2005b}. In this attack, Eve
intercepts and performs a complete von Neumann measurement on each
quantum state sent out by Alice. She then generates a new
time-shifted signal based on her measurement result. In the extreme
case where there is complete detector efficiency mismatch (that is
to say, there is a time window where the detector for the bit ``0''
is active while the detector for the bit ``1'' is completely
inactive, and vice versa), Eve can acquire full information on the
final key without introducing any error.

In this paper, we propose a simple attack that exploits the same
imperfection. More specifically, in our attack, Eve does not measure
the quantum state sent by Alice. Instead, Eve simply time-shifts
each of Alice's signal forward or backward as she wishes. It turns
out that the security of some practical QKD systems, particularly
those with time-multiplexed SPDs, could be totally compromised by
this simple attack.
We emphasize that our attack can be implemented with current technology.

This paper is organized as follows: In Section $2$, we
summarize the basic results in \cite{Makarov2005b} and then propose
our new eavesdropping strategy. This is followed by a comparison of
these two attacks. In Section $3$, we discuss the security loophole
of QKD systems with a time-multiplexed SPD. Finally, in Section $4$,
we discuss some counter measures against our and earlier attacks.

\section{Eavesdropping strategies exploiting detector efficiency mismatch\label{strategy}}

In a BB84 QKD system \cite{Bennett1984}, typically, Bob uses two
separate SPDs, which are labeled as SPD0 and SPD1, to
detect bit ``0'' and bit ``1'' respectively. As we discussed in
Section $1$, to minimize the dark count rate, each of these two SPDs
is only activated for a narrow time window when the signal is
expected. Owing to the different responses of the two APDs and other
imperfections in electronics, the time-dependent efficiencies of
these two SPDs are not identical. Because the width of SPD's open
window (a few ns) is often substantially larger, or at least not
shorter than the laser pulse duration (a few hundreds ps), Alice and
Bob can synchronize the laser pulse with the center of SPD's open
window. This ensures that the small detector efficiency mismatch
will not affect the normal operation of the QKD system.

\begin{figure}
\centerline{\epsfig{file=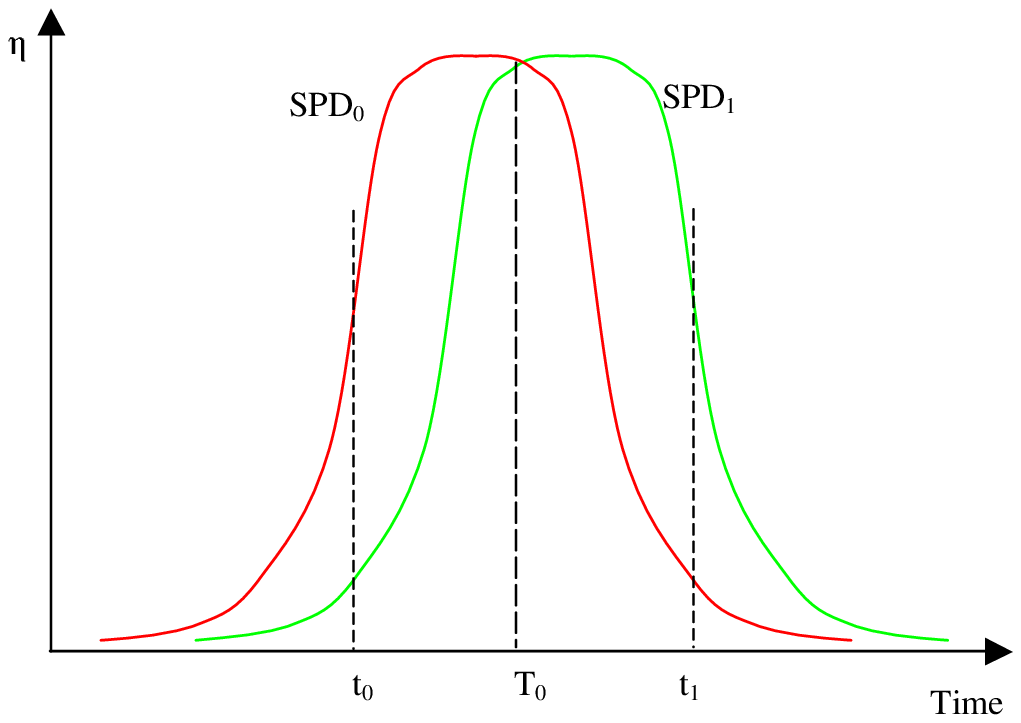, width=8.2cm}} 
\fcaption{\label{fig:SPD-a}
The time-dependence efficiencies of SPDs.
The dark (light) curve corresponds to the efficiency of SPD0 (SPD1).
At time $t_0$, SPD0 is more sensitive to the incoming photon than SPD1.
}
\end{figure}
Ref.~\cite{Makarov2005b} considers the time-dependent detector
efficiency of the two detectors. Fig.\ref{fig:SPD-a} shows a diagram
of the time-dependent efficiency of the two SPDs: at time $t_0$, the
efficiency of SPD0, $\eta_0(t_0)$, is much higher than that of SPD1,
$\eta_1(t_0)$, while at time $t_1$, $\eta_0(t_1) \ll \eta_1(t_1)$.
Taking a symmetrical assumption \cite{Makarov2005b}, we can define
\begin{equation}
\myeta = \frac{\eta_1(t_0)}{\eta_0(t_0)} = \frac{\eta_0(t_1)}{\eta_1(t_1)}
\end{equation}
where $\myeta \in [0,1]$. In the extreme case favorable to Eve,
$\myeta=0$, which means only SPD0 is active in time $t_0$, and only
SPD1 is active at time $t_1$.

The intercept-resend attack described in \cite{Makarov2005b} goes as
follows:
\begin{enumerate}
\item[(1)]
Eve intercepts quantum states from Alice and measures each of them in a randomly chosen basis;
\item[(2)]
According to her measurement result, Eve prepares a new quantum
state (faked    state \cite{Makarov2005a}) in a different basis with
a different bit value. For example, if she     measures in Z basis
and gets bit ``0'' (labeled as $Z_0$), then she prepares bit ``1''
in X basis ($X_1$);
\item[(3)]
According to her measurement result, Eve sends out her faked state
at different time so that it arrives at Bob's SPD at either time
$t_0$ (corresponding to her measurement result ``0'') or $t_1$
(corresponding to her   measurement result ``1'').
\end{enumerate}

Assuming the intrinsic QBER in Alice and Bob's system is zero, the
QBER caused by this attack can be derived as \cite{Makarov2005b}:
\begin{equation}
QBER = \frac{2 \eta_0(t_1) + 2 \eta_1(t_0)}{\eta_0(t_0)+3\eta_0(t_1)+3\eta_1(t_0)+\eta_1(t_1)}
\end{equation}

In one extreme case when $\myeta=0$ (which corresponds to $\eta_0(t_1)
=  \eta_1(t_0) =0$), QBER=0. So Eve can acquire full information on
the sifted key without introducing any errors. In the other extreme
case, when $\myeta=1$ (the two SPDs are perfectly matched),
Eve causes QBER=$1/2$. So, Eve is only doing a random guess. This is worse than
the standard intercept-resend attack (which causes QBER =$1/4$, and
requires exactly the same equipment capabilities from Eve). The
attack \cite{Makarov2005b} becomes more efficient than the standard
intercept-resend attack for $\myeta<0.2$. In the intermediate case
when $0<\myeta<0.2$, the QBER calculated from (2) could be lower than
the proven lower bound of standard BB84 system. In other words, a
direct application of standard security proofs, e.g. GLLP
\cite{Gottesman2004}, without taking into account of detector
efficiency mismatch, is invalid.

Note that, to implement the attack in \cite{Makarov2005b}, Eve will
need a complicated detection (similar to Bob's system) and resend
(similar to Alice's system) system. If we assume that Eve builds her
``practical'' eavesdropping device based on today's technology, she
will also experience the problem of low detection efficiency and
will introduce additional errors (even in the case of $\myeta=0$) due
to the imperfections in her setup.

\begin{figure}
\centerline{\epsfig{file=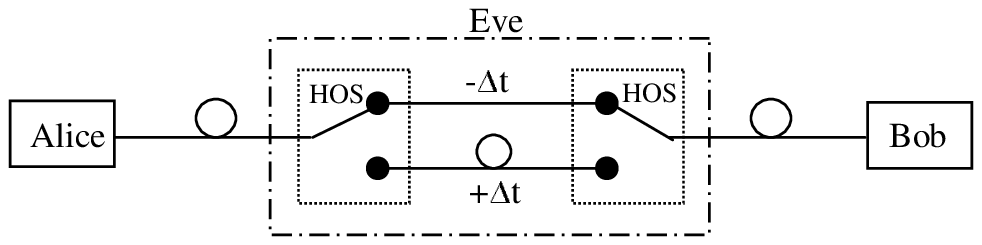, width=8.2cm}} 
\fcaption{\label{fig:attack1}
A schematic diagram of Eve's attack. HOS: high-speed optical switch.
}
\end{figure}

Here we propose a simple and practical attack that exploits the same
imperfection. Fig.\ref{fig:attack1} shows the schematic diagram of
the experimental system that implements our attack.
Instead of measuring Alice's quantum state, Eve just randomly shifts
the time of Alice's quantum state to make sure that it arrives at
Bob's detector at either time $t_0$ or at time $t_1$. If Eve chooses
time $t_0$ ( $t_1$ ), whenever Bob detects a signal, with the
probability of $1/(\myeta+1)$, the bit value will be ``0'' ( ``1'' ).
[Here, we assume equal prior probability for the bit ``0'' and ``1''
emitted by Alice.] In the extreme case when $\myeta=0$, this
probability is equal to one. Because the probability that Eve
guesses Bob's bit value wrong is $\myeta/(\myeta+1)$, So Eve's knowledge
about the final key is given by:
\begin{equation}
I(B:E) = 1 - h(\myeta/(\myeta+1)),
\end{equation}
where $h(x)=-x \log_2(x)-(1-x) \log_2(1-x)$ is the binary Shannon
entropy function.
Note that in this attack, Eve does not measure
Alice's state. Therefore, Eve never introduces any errors.

We now show that whenever $I(B:E)>0$,
the final key shared between Alice and Bob is insecure if they are not aware of Eve's attack.
In order to show this, we compute an upper bound on the key generation rate under Eve's attack.
One upper bound is the conditional mutual information \cite{Maurer1999} given by
\begin{equation}
I(A:B|E) = h(\myeta/(\myeta+1)) .
\end{equation}
Thus,
if Alice and Bob are not aware of our attack, then they
would generate keys at a rate equal to $1$
since there are no bit errors\footnote{In reality, due to the finiteness of the number of signals,
Alice and Bob may choose the key generation rate to be $1-h(\epsilon)$, where $\epsilon$ is some security parameter.}.
Therefore, whenever $h(\myeta/(\myeta+1))<1$, or in other words, when Alice and Bob generate keys at a rate higher than the upper bound,
the final key shared between them is insecure.
To quantify this key rate upper bound, we may substitute in an experimental value of $\myeta=2$ observed in our experiment.
This gives us an upper bound of $I(A:B|E) = h(2/3) = 0.9183$.
Thus, our attack does cause a moderate decrease in the key rate upper bound in this case.


Comparing with the attack in \cite{Makarov2005b}, our attack is
simpler and can be easily realized with today's technology: Eve can
use high speed optical switches to re-route Alice's signal through
either a long optical path or a short optical path to achieve the
desired time shift. Another advantage of our attack is that Eve
never introduces any errors. So, it is hard for Alice and Bob to
detect Eve's presence.

One drawback of our attack (from the point of view of Eve) is that
Eve can't get the full information (unless $\myeta=0$) on the sifted
key. So, it might still be possible for Alice and Bob to work out a
secure key, provided that they can correctly upper bound Eve's
information.

In both the attack proposed in \cite{Makarov2005b} and the one
proposed here, Bob's detection efficiency is lower than normal. In
the previous case, Eve can compensate this decrease by using a
stronger faked state, or by using a low-loss quantum channel, or by
placing her intercept unit near Alice and resend unit near Bob to
bypass the quantum channel\cite{Makarov2005b}, whereas in our
attack, Eve has to use a low-loss channel.
Another alternative to using a low-loss channel is for Eve to mask herself by introducing additional loss during the calibration phase.
This can be done with current technology.

In next section, we will show that for some QKD design, where Bob
uses the same SPD to detect both bit ``0'' and bit ``1'', Eve could
time-shift both the synchronization signal and quantum signal to
acquire full information without introducing any errors. In other
words, our attack is fatal there.

\section{Security loophole of QKD system with a time-multiplexed SPD\label{time-mux}}

In a standard design of BB84 QKD system, Alice sends out
her quantum bit in a randomly chosen basis, while Bob also randomly
chooses his measurement basis and uses two separate SPDs to detect
bit ``0'' or bit ``1''. In a modified system design
\cite{Marand1995,Makarov2004}, Bob detects both bit ``0'' and bit
``1'' with the same SPD, by employing time-multiplexing technique.

\begin{figure}
\centerline{\epsfig{file=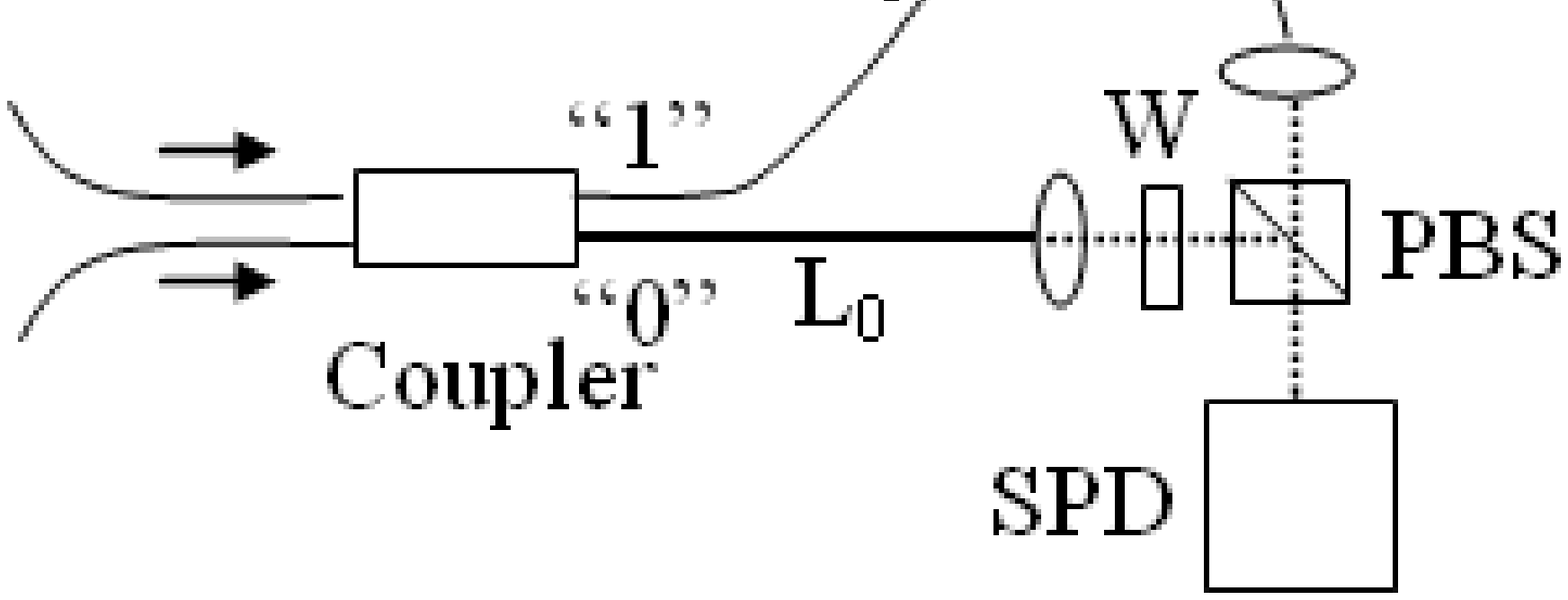, width=8cm}} 
\fcaption{\label{fig:setup1} A scheme for detecting both bit ``0''
and bit ``1'' with the same SPD: W--wave plate, PBS--polarization
beam splitter. }
\end{figure}
In this design, the expected arrival time $t_0$ of bit ``0'' (to
SPD) is different from the expected arrival time $t_1$ of bit ``1'',
with the time difference
\begin{equation} \Delta t = t_1-t_0
\end{equation}
determined by the optical length difference $n(L_1-L_0)$, as shown
in Fig.~\ref{fig:setup1}. Here, $n$ is the effective refractive
index of the fibers and we assume $t_1 > t_0$. So, it's possible for
Bob to distinguish bit ``0'' and bit ``1'' from the detection time
information.

Besides the advantage of saving one SPD (which is a most expensive
component in a QKD system), this specific design also helps to
minimize the asymmetry between the detection efficiency of bit ``0''
and bit ``1''.   Unfortunately, as we will demonstrate below, this
design may also open a loophole that will allow Eve to launch the
``time-shift'' attack that we discussed in Section $2$.

Note that if Bob's SPD works in a gated mode (as the case for most
Telecommunication wavelength QKD), it has to be activated twice for
each incoming pulse (at time $t_0$ and $t_1$). Since there is no
overlap between these two open time windows, this is equivalent to
the case $\myeta=0$ in Section $2$. Therefore, Eve could acquire full
information on the sifted key without introducing any error. Before
we go into the details of this attack, we will first discuss the
synchronization of Alice's laser pulse with Bob's detection window
in a practical QKD system. This synchronization issue is crucial for
our discussion.

\begin{figure}
\centerline{\epsfig{file=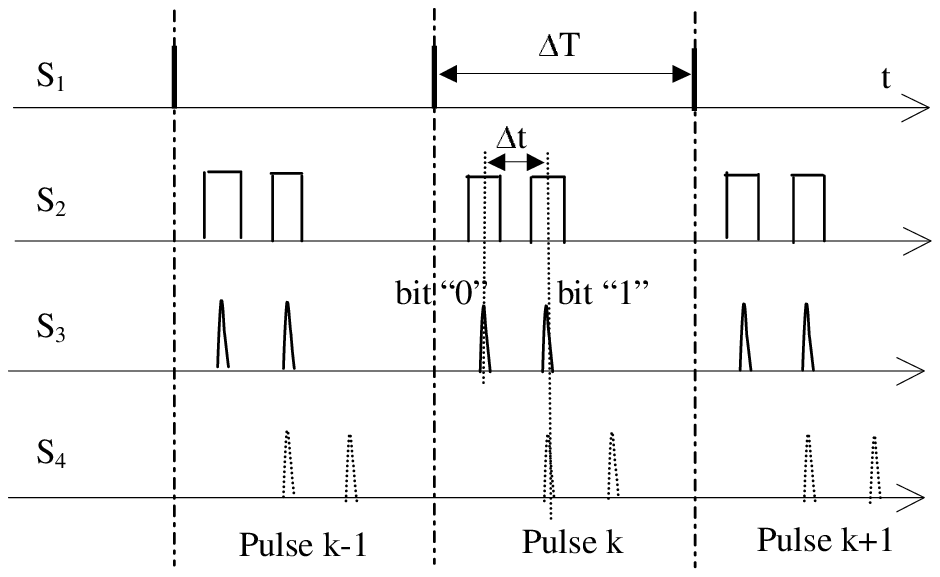, width=8.2cm}} 
\fcaption{\label{fig:sync} Synchronization between Alice's pulse and
Bob's SPDs: $S_1$-classical Sync pulses, $S_2$-SPD's gating signals,
$S_3$-signal pulses without Eve, $S_4$-Eve shifts the signal pulses
by $+\Delta t$.}
\end{figure}
A widely used synchronization method in a practical QKD system is by
multiplexing a strong synchronizing laser pulse with the weak signal
pulse by employing wavelength division multiplexing (WDM) technique
\cite{Hughes2000,Yuan2005}. In \cite{Yuan2005}, Alice multiplexes
each weak signal pulse (1.55 $\mu$m) with a strong timing pulse
(1.3$\mu$m) and sends them to Bob through the same optical fiber. On
the receiver's side, Bob first separates the timing pulse from the
signal pulse with a WDM component, and then detects the timing pulse with an APD,
which in turn produces two electrical gating signals to control the
two SPDs (or in the case of time-multiplexed SPD,  these two
electrical gating signals activate the SPD twice for each signal
pulse). The time relations of various signals are depicted in
Fig.~\ref{fig:sync}.

Eve's eavesdropping strategy is quite similar to the one we proposed
in Section $2$: Eve doesn't measure Alice's quantum state. Instead,
she flips the bit value first (for phase encoding BB84 protocol,
this can easily be done by using a phase modulator to introduce an
additional $\pi$ phase shift between the signal and reference pulses
sent by Alice) and then randomly time-shifts the pulse (relative to
the strong sync pulse) by an amount of either $-\Delta t$ or
$+\Delta t$. Fig.~\ref{fig:sync} ($S_4$) shows the situation when
Eve does the $+\Delta t$ shift. Obviously, in this case, Bob can
only detect bit ``1'' or nothing; the bit ``0'' is blinded.
Similarly, when Eve does the ``$-\Delta t$'' shift, Bob can only
detect bit ``0'' or nothing. So, whenever Bob has a detection event,
Eve has full information about his bit value.

\begin{table}
\tcaption{\label{table-attack1}Alice sends out $Z_0$ and Bob measures in Z basis.
}
\centerline{
\begin{tabular}{|c|c|c|c|c|c|}
\hline
{\footnotesize Quantum state } & {\footnotesize Eve flips the}  & \multicolumn{2}{c|}{\footnotesize Eve does random time shift} & \multicolumn{2}{c|}{\footnotesize Bob's measurement results} \\
{\footnotesize from Alice} & {\footnotesize bit value} & \multicolumn{2}{c|}{}& \multicolumn{2}{c|}{\footnotesize in Z basis} \\
& & {\footnotesize Time shift} & {\footnotesize Possibility} & {\footnotesize Bit value} & {\footnotesize Possibility} \\ \hline
\multirow{4}{*}{$Z_0$} & \multirow{4}{*}{$Z_1$} & \multirow{2}{*}{$-\Delta t$} & \multirow{2}{*}{$\frac{1}{2}$} & 0 & $\frac{1}{2}$ \\
& & & & 1 & 0 \\
& & \multirow{2}{*}{$+\Delta t$} & \multirow{2}{*}{$\frac{1}{2}$} & 0 & 0 \\
& & & & 1 & 0 \\ \hline
\end{tabular}
}
\end{table}
Table~\ref{table-attack1}  summarizes all possible events when Alice
sends out $Z_0$ and Bob measures in Z basis. The other three cases
can be easily worked out by symmetry. Again, as we explained in
Section $2$, Eve's attack does not introduce any error.

Note that if the time interval $\Delta t$ between the two open
windows of Bob' SPD is exactly equal to half of the laser pulse
period $\Delta T$, as the case in \cite{Makarov2004}, this attack
cannot be applied\cite{error}. The reason is the following: Owing to
Eve's shifting of the pulse, the arriving time of the ``blinded''
bit will overlap with one of the SPD's two open windows for the
successive signal. For example, suppose Eve does ``$+\Delta t$''
shift to the $k^{th}$ laser pulse, then with $50\%$ chance, the
incoming photon will be detected at the time corresponding to bit
``0'' for the $(k+1)^{th}$ pulse.  It's easy to see that QBER in
this case is $25\%$.

Note that QKD systems with a long SPD gating period (larger than
$3\Delta t$) may be immune to this kind of attack. In this case, Bob
may monitor counts fall outside the predetermined ``0'' and
``1''windows. For example, in \cite{Townsend}, the SPD is gated with
$\sim 100$ns pulse, while the time interval $\Delta t$ between bit
``0'' and bit ``1'' is a $6.2$ns. In this case, the SPD is only
activated once for each incoming signal, and Bob can distinguish bit
``0'' from bit ``1'' by measuring the time difference between the
sync pulse and the detected photon.

\section{Discussion\label{discussion}}

The recently proposed eavesdropping strategy exploits the detector
efficiency mismatch of SPDs\cite{Makarov2005b}. In this paper, we
propose a simple time-shifting attack that exploits the same
imperfection.
We emphasize that our attack is simple and feasible with current technology.
In particular, we demonstrate a QKD system, where Bob
employs time-multiplexing technique to detect both bit ``0'' and bit
``1'' with the same SPD, is especially vulnerable to this kind of
attack.

People may ask that even there is a mismatch between the efficiencies
of two SPDs, how Eve can know that. We remark that in quantum
cryptography, if we cannot prove that Eve is incapable of getting this
information, then we have to assume that this information is available
to Eve.
In fact, in this specific case, Eve, in principle, does have a way to acquire
this information during the normal quantum key distribution process: she
can randomly block a small fraction of the pulses from Alice and resend
{\it faked pulses} to Bob (she only changes a {\it small} fraction, so the QBER
does not change significantly). Each faked pulse is randomly prepared in
one of the four states as in the BB84 protocol, and its delay time can be
tuned by Eve.
After the classical communication stage, Eve knows
Bob's basis choice for each detected faked pulse.
In the case when
they (Eve and Bob) use the same basis, Eve knows for sure which SPD
clicks.
Let us assume that the properties of the two SPDs stay constant over
a long period of time during an experiment. In this case, given a long
enough
time, Eve can simply repeat her attacks and thus gather enough statistical
data to
determine the efficiencies of
both SPDs at various delay times.
With this information, she can launch
the time-shift attack we discussed in this paper.
Another scenario where Eve can obtain the efficiency information is that Eve is actually
the producer of the QKD system Alice and Bob use.
Thus, in principle, Eve knows the efficiency
mismatch between the two SPDs.

To counter Eve's attack, Alice and Bob could develop various counter
measures, such as those discussed in \cite{Makarov2005b}. Other
counter measures include: strict checks for detection rate and loss
in the quantum channel; applying phase shift setting to Bob's phase
modulator for a narrower time period than [$-\Delta t$, $+\Delta t$]
relative to the normal pulse position; randomly shifting the gating
window; and using a non-gating SPD.

We note that a recently proposed single SPD QKD system is also
immune to this attack \cite{LaGasse2005}. In a phase encoding BB84
version of this design, instead of randomly selecting from a set of
two values, Bob's phase modulation is randomly selected from a set
of four values, which is identical to the set for Alice's phase
modulation. In this case, Bob not only randomly chooses his
measuring basis for each incoming pulse, he also randomly determines
which SPD (or which time window in the case of time-multiplexed SPD)
is used for detecting bit ``0'' or bit ``1''. Bob broadcasts his
basis choice, but keeps his choice of detector (for the bit ``0'' or
``1'') secret. In such a set-up, even if Eve has the information
about which detector clicks, Eve still cannot work out Bob's bit
value because she does not know which detector corresponds to bit
``0''.
On another hand,
the set-up in \cite{LaGasse2005} may open up an unrelated security loophole.
More concretely,
Eve may try to read out Bob's detector
assignments by employing a strong external laser pulse, as described
in \cite{Largepulse}.

We conclude with some general comments on the security of QKD.
First, a
security proof is only as good as its underlying assumptions. Once a
security loophole has been discovered, it is often not too difficult
to develop counter-measures that will plug the loophole and regain
proofs of unconditional security of QKD. One example is the PNS
attack that we mentioned in Section $1$. However, the hard part is
how to find out those security loopholes in the first place. A QKD
system is a complicated system with many intrinsic imperfections. It
is, thus, very important to conduct extensive research in those
imperfections carefully to see if they are innocent or fatal for
security. We need more quantum hackers in the field. The
investigation of loopholes and counter-measures in practical QKD
systems plays a complementary role to security proofs.

Second, implementing a countermeasure for some security loophole may open up new loopholes (such as the one mentioned above).
Also, regarding countermeasures,
in our opinion, there is a big difference between what Bob could do in principle and what Bob actually does in practice.

Third, it is important to construct a list of potential countermeasures, and to compare and contrast them.
This task is specified in the ARDA's quantum cryptography roadmap \cite{arda} as one of the major milestones.
Such a study will help us to better understand the pros and cons of the various countermeasures and eventually lead to the identification of the best countermeasure in practice.

Fourth, given that a practical QKD system will always have imperfections,
one might wonder if QKD systems offer any real advantage over
conventional systems. Our answer is three-fold. First of all,
implementation loopholes are a fact of life. Even conventional
security systems such as smart cards suffer implementation
loopholes. For instance, Eve may attempt to read off a private key
from a smart card by using various techniques (including X-ray) to
reverse-engineer the circuit embedded in a smart card. Second, QKD
can be used in concatenation with a conventional system to ensure
security\cite{concatenation}. By defending in depth, QKD can only
increase security, not reduce it. Third, QKD has an important
advantage of being future-proof: The signals are quantum. Once the
transmission is done, there is no transcript for the transmission.
For an eavesdropper to launch a quantum attack, she has to possess
much of the quantum technology during the quantum transmission. In
contrast, in a standard Diffie-Hellman public-key key exchange
scheme, Eve has a complete transcript of the transmission and can
save such a transcript for decades to wait for unexpected future
advances in hardware and algorithms. Given that public key
crypto-systems was itself an unexpected discovery made only three
decades ago, our view is that it will be complacent to believe that
our standard public key crypto-systems will be safe forever.
Therefore, it pays to diversify one's risk by defending in depth
with a QKD system in concatenation with a conventional cryptosystem.

\paragraph{Acknowledgments}
The authors are grateful to the anonymous reviewers for pointing out
one error in the first version and making numerous important
comments. Financial support from NSERC, CIAR, CRC Program, CFI, OIT,
PREA and CIPI are gratefully acknowledged.


\begin{thebibliography}{00}
\bibitem{error}
In the first version, we proposed an attack on this type QKD system
by tampering both the synchronization pulse and signal pulse. As
pointed out by an anonymous reviewer, in this case, the time-shifted
pulse may be modulated by Bob's phase modulator
incorrectly.
\bibitem{concatenation}
For example, Alice/Bob may produce one key with QKD system and
another key with conventional cryptosystem. They can produce the
final key by doing a ``XOR'' operation. Eve has to break both the two
systems to get the final key.
\end{thebibliography}

\paragraph{References}

\end{document}